\documentclass[runningheads,a4paper]{llncs}

\usepackage{amssymb}
\usepackage{graphicx}
\usepackage{verbatim}

\usepackage{url}

\newcommand{\keywords}[1]{\par\addvspace\baselineskip
\noindent\keywordname\enspace\ignorespaces#1}

\begin{document}

\mainmatter  % start of an individual contribution

\title{Rfuzzy framework}

% \titlerunning{Rfuzzy framework}

\author{Victor Pablos Ceruelo \and Susana Munoz-Hernandez \and Hannes Strass}
%\authorrunning{Victor Pablos Ceruelo \and Susana Munoz-Hernandez \and Hannes Strass}

\institute{Universidad Polit{\'e}cnica de Madrid
\footnote{This work is partially supported by the project DESAFIOS - TIN 2006-15660-C02-02 from the Spanish Ministry of Education and Science and project PROMESAS -  S-0505/TIC/0407 from the Madrid Regional Government.}  \\
\email{\{vpablos,susana\} at fi.upm.es, hannes.strass at alumnos.upm.es} \\
\url{http://babel.ls.fi.upm.es/}
}

\maketitle

\setcounter{page}{62}

\begin{abstract}
  Fuzzy reasoning is a very productive research field that during the
  last years has provided a number of theoretical approaches and
  practical implementation prototypes. Nevertheless, the classical
  implementations, like Fril, are not adapted to the latest formal
  approaches, like multi-adjoint logic semantics.

  Some promising implementations, like Fuzzy Prolog, are so general
  that the regular user/programmer does not feel comfortable because
  either representation of fuzzy concepts is complex or the results
  difficult to interpret.

  In this paper we present a modern framework, \emph{Rfuzzy}, that is
  modelling multi-adjoint logic. It provides some extensions as default
  values (to represent missing information, even partial default
  values) and typed variables. Rfuzzy represents the truth value of
  predicates through facts, rules and functions.  Rfuzzy answers
  queries with direct results (instead of constraints) and it is easy
  to use for any person that wants to represent a problem using fuzzy
  reasoning in a simple way (by using the classical representation
  with real numbers).

\end{abstract}

%%%%%%%%%%%%%%%%%%%%%%%%%%%%%%%%%%%%%%%%%%%%%%%%%%%%%%%%%%%%%%%%%%%
%%%%%%%%%%%%    INTRODUCTION     %%%%%%%%%%%%%%%%%%%%%%%%%%%%%%%%%%
%%%%%%%%%%%%%%%%%%%%%%%%%%%%%%%%%%%%%%%%%%%%%%%%%%%%%%%%%%%%%%%%%%%

\section{Introduction}
\label{sec:intro}

One of the reasoning models that is more useful to represent real
situations is fuzzy reasoning. Indeed, world information is not
represented in a crisp way. Its representation is imperfect, fuzzy,
etc., so that the management of uncertainty is very important in
knowledge representation. 
There are multiple frameworks for incorporating 
uncertainty in logic programming:

\begin{itemize}
     \item fuzzy set theory \cite{Cao00,Shapiro83,Emden86},
     \item probability theory
     \cite{Fuhr00,LakshmananShiri94,Lukasiewicz01,Ng91,Ng93},
     \item multi-valued logic
     \cite{Fitting91,Kifer88,Kifer92,Lakshmanan94,Lakshmanan01,Subrahmanian87},
     \item possibilistic logic \cite{Dubois91,Wagner97,Wagner98}
\end{itemize}

Despite of the multitude of theoretical approaches to this issue, 
few of them resulted in actual practically usable tools.
Since Logic Programming is traditionally used in 
Knowledge Representation and Reasoning, 
we argue it is perfectly well-suited to implement 
a fuzzy reasoning tool as ours.

%%%%%%%%%%%%    FUZZY APPROACHES   %%%%%%%%%%%%%%%%%%%%%%%%%%%%%%%%
\subsection{Fuzzy Approaches in Logic Programming}
\label{sec:approaches}

The result of introducing Fuzzy Logic into Logic Programming has been
the development of several fuzzy systems over Prolog. These systems
replace the inference mechanism of Prolog, SLD-resolution, with a
fuzzy variant that is able to handle partial truth. Most of these
systems implement the fuzzy resolution introduced by Lee in
\cite{lee1}, as 
the Prolog-Elf system \cite{ishizuka85prologelf}, 
the FRIL Prolog system \cite{fril} and 
the F-Prolog language \cite{li90fuzzy}. 
However, there is no common method for fuzzifying Prolog, 
as has been noted in \cite{Shen}. 

Some of these Fuzzy Prolog systems only consider the predicates' 
fuzziness whereas other systems consider fuzzy facts or fuzzy rules. 
There is no agreement about which fuzzy logic should be used. 
Most of them use min-max logic 
(for modelling the conjunction and disjunction operations) 
but other systems just use Lukasiewicz logic
\cite{klawonn94lukasiewicz}.

Furthermore, logic programming is considered a useful tool for
implementing methods for reasoning with uncertainty in
\cite{Shapiro83}.

There is also an extension of constraint logic programming
\cite{Bistarelli01}, which can model logics based on semiring
structures. This framework can model min-max fuzzy logic, which is the
only logic with semiring structure.
% TODO: missing justification (a citation in the easiest of cases)

Another theoretical model for fuzzy logic programming without negation
has been proposed by Vojtas in \cite{Vojt1}, which deals with
many-valued implications.

%%%%%%%%%%%%    FUZZY PROLOG   %%%%%%%%%%%%%%%%%%%%%%%%%%%%%%%%%%%%
\subsection{Fuzzy Prolog}
\label{subsec:fuzzy-prolog}

One of the most promising fuzzy tools for Prolog was the ``Fuzzy
Prolog'' system
\cite{Vaucheret_LPAR02,Susana_FSS04}. 
% Vaucheret_ICLP02, <-- Refs [28] and [29] have the same title. 
% It seems that you presented a poster in ICLP and a long paper in
% LPAR. Please remove [28]
%
The most important advantages against the other approaches are:
\begin{enumerate}

\item A truth value will be a finite union of sub-intervals on
  $[0,1]$. An interval is a particular case of union of one element,
  and a unique truth value is a particular case of having an interval
  with only one element.

\item A truth value is propagated through the rules by means of
  an \emph{aggregation operator}. The definition of this 
  \emph{aggregation operator} is general and 
  it subsumes conjunctive operators (triangular
  norms \cite{Norms00} like min, prod, etc.), disjunctive operators
  \cite{Tri_Cub_Cas} (triangular co-norms, like max, sum, etc.),
  average operators (averages as arithmetic average, quasi-linear
  average, etc) and hybrid operators (combinations of the above
  operators \cite{Prad_Tri_Cal}). 

\item Crisp and fuzzy reasoning are consistently combined 
  \cite{Susana_fuzzyneg_AGP02}.
\end{enumerate}

Fuzzy Prolog adds fuzziness to a Prolog compiler using CLP(${\cal R}$)
instead of implementing a new fuzzy resolution, as other former fuzzy
Prologs do. 
It represents intervals as constraints over real numbers and 
\emph{aggregation operators} as operations with these constraints,
so it uses Prolog's built-in inference mechanism 
to handle the concept of partial truth. 

%% There are other proposals, e.g. in %\cite{nguyen1,klir1},
%% \cite{klir1}, that provide
%% an interpretation of truth values as intervals, but Fuzzy Prolog 
%% proposed for the first time to generalise this concept to union of
%% intervals.

%%%%%%%%%%%%%%%%%%%%%%%%%%%%%%%%%%%%%%%%%%%%%%%%%%%%%%%%%%%%%%%%%%%
%%%%%%%%%%%%    MOTIVATION AND   %%%%%%%%%%%%%%%%%%%%%%%%%%%%%%%%%%
%%%%%%%%%%%%    RFUZZY APPROACH  %%%%%%%%%%%%%%%%%%%%%%%%%%%%%%%%%%
%%%%%%%%%%%%%%%%%%%%%%%%%%%%%%%%%%%%%%%%%%%%%%%%%%%%%%%%%%%%%%%%%%%

\subsection{Motivation and RFuzzy Approach}
\label{subsec:motivation}

Over the last few years several papers have been published by Medina
et al.
(\cite{M_Adjoint_Continuous,M_Adjoint_Procedural,M_Adjoint_Completeness})
about multi-adjoint programming, which describe a theoretical model,
but no means of serious implementations apart from promising
prototypes \cite{Abietar07}.

Indeed, that was the reason for developing Fuzzy Prolog \cite{Susana_FSS04}.
Fuzzy Prolog  is a very expressive tool which allows the 
user \footnote{We refer as 'user' to the programmer that
  wants to represent a fuzzy problem in a programming framework to make
  queries and obtain results.}
to program almost everything, but we have to pay for this
expressiveness.
The cost is a complex syntax difficult to understand.

The motivation for developing Rfuzzy is mainly focused on reducing this 
complex syntax. This reduction is based on three ideas:
\begin{enumerate}
\item Use real numbers instead of intervals between real numbers to 
  represent truth values.
  Fuzzy Prolog answers to user queries are intervals like \\
  {\it it\_will\_rain\ (tonight, [0,\ 1])}. 
  This is a bit difficult to understand by normal users.
  Truth value of this example is between 0 and 1, and this means that 
  program can not conclude anything about the predicate truth value.
\item Whenever it is possible, do not answer user queries using 
  constraints.
  A Fuzzy Prolog answer to an user query can be a constraint, like \\
  {\it it\_will\_rain\ (tomorrow, [X,\ Y]), X $>$ 0, X $<$ 1, Y $>$ 0, Y $<$ 1}.
  The meaning of this example is exactly the same as the meaning of 
  the previous one, but it is slightly more difficult to understand it.
\item Truth value variables do not need to be coded. 
  Taking care of variables to manage the predicates truth value 
  introduces errors and makes the code illegible, without giving us 
  any advantage.
\end{enumerate}

Rfuzzy uses real numbers to represent truth values and its replies are 
never constraints. 
Besides, it is able to distinguish between crisp and fuzzy predicates
and it manages the introduction of truth value variables, so the 
user does not need to take care of them.
Truth variables are always introduced at the end of 
the predicate arguments list, so it can be seen as some syntactic 
sugar. We explain this in subsection \ref{doing-queries-with-truth-values}.

From the point of view of expressiveness, we can remark that RFuzzy offers
to the user the ability to define types, general and conditioned
default values and truth value representations by means of facts,
functions or rules. 
Besides, it implements multi-adjoint logic with
representation of the concept of credibility of the rules, so %TODO: remarks about the semantics of credibility
it is one of the first tools that are actually modelling multi-adjoint logic
% \footnote{For more details about the semantics consult 
% \url{http://babel.ls.fi.upm.es/software/rfuzzy/}. 
% A paper (pending of acceptance in an international conference) 
% that contains a complete formalization of the semantics of RFuzzy
% can be found there.}.
\footnote{A complete formalization of the semantics of RFuzzy with a
  description of a least model semantics, a least fixpoint semantics,
  an operational semantics and the prove of their equivalence 
  can be downloaded at 
  \url{http://babel.ls.fi.upm.es/software/rfuzzy/}.
  This paper has been submitted and is pending of
  acceptance in an international conference.}.
 
%%%%%%%%%%%%%%%%%%%%%%%%%%%%%%%%%%%%%%%%%%%%%%%%%%%%%%%%%%%%%%%%%%%
%%%%%%%%%%%%    RFUZZY SYNTAX   %%%%%%%%%%%%%%%%%%%%%%%%%%%%%%%%%%%
%%%%%%%%%%%%%%%%%%%%%%%%%%%%%%%%%%%%%%%%%%%%%%%%%%%%%%%%%%%%%%%%%%%

\section{Rfuzzy syntax}
\label{rfuzzy:syntax}

In this section we are going to describe RFuzzy's syntax.
Rfuzzy defines the syntax of a new subset of Prolog predicates 
to work with truth values and to assign credibility to rules. 
The extensions that we have added to provide fuzziness of 
predicates are:
type information, truth values (for facts, functions and rules) 
and default truth values.
 
RFuzzy shares with Fuzzy Prolog most of its nice
expressive characteristics: Prolog-like syntax (based on using facts
and clauses), use of any aggregation operator, flexibility of query
syntax, constructivity of the answers, etc. Nevertheless, RFuzzy is
simpler than Fuzzy Prolog for the user because the truth values are
simple real numbers instead of the general structures of Fuzzy Prolog.

%%%%%%%%%%%%    TYPE DEFINITION    %%%%%%%%%%%%%%%%%%%%%%%%%%%%%%%%%%%%%%
\subsection{Type definition}
\label{type-definition}

Prolog does not have types. 
Prolog code are formulas and at execution time it looks for all of them to 
be true. 
To do that, it generates a Herbrand Universe and tries to 
substitute every variable with a Herbrand term. 
As we do not want programs to look for an answer infinitely, 
we offer the user a facility to restrict the set of possible solutions.
This extension is named ``types'' and its syntax is shown 
in (\ref{eq-def-type}).
%
% \textbf{=>}
\begin{equation}
\textbf{:- set\_prop}\ pred/ar\ \textbf{=\textgreater}\ type\_pred\_1/1\ [,\ type\_pred\_2/1\ ]^*\ .
\label{eq-def-type}
\end{equation}
where {\it set\_prop} is a reserved word, {\it pred} is the name
of the typed predicate, {\it ar} is its arity and
{\it type\_pred\_\{n\}} is the predicate used to assure that the
value given to the argument in the position {\it n} of a call to
{\it pred/ar} is correctly typed. 
Predicate {\it type\_pred\_\{n\}} must have arity 1.  
The example below shows that the two arguments of the predicate
{\it has\_lower\_price/2} have to be of type {\it car/1} and
which individuals belong to that type.

\begin{verbatim}
:- set_prop has_lower_price/2 => car/1, car/1.
car(vw_caddy).
car(alfa_romeo_gt).
car(aston_martin_bulldog).
car(lamborghini_urraco).
\end{verbatim}

%%%%%%%%%%%%   FACT TRUTH VALUE %%%%%%%%%%%%%%%%%%%%%%%%%%%%%%%%%%
\subsection{Fact truth value}
\label{fact-truth-value}

Fuzzy facts are facts to which we assign a truth value. 
To code them in programs we offer a special syntax, so Prolog 
can distinguish between normal facts and fuzzy facts.
This syntax is shown in (\ref{eq-def-fact}).
\begin{equation}
pred(args)\ \textbf{value}\ truth\_val.
\label{eq-def-fact}
\end{equation}
 
%To define a truth value for a predicate when applied to an individual, 
Arguments ( {\it args} ) should be ground and the truth value ( {\it
truth\_val} ) must be a real number between 0 and 1.  The example 
below defines that the car {\it
alfa\_romeo\_gt} is an {\it expensive\_car} with a truth value 0.6.

\begin{verbatim}
expensive_car(alfa_romeo_gt) value 0.6 .
\end{verbatim}

%%%%%%%%%%%%    FUNCTIONAL TRUTH VALUE %%%%%%%%%%%%%%%%%%%%%%%%%%%
\subsection{Functional truth value}
\label{functional-truth-value}

Fact truth value definition (see subsection \ref{fact-truth-value}) 
is worth for a finite (and relative small) number of individuals.
As we may want to define a big amount of individuals, we need
more than this.

\begin{figure}
\centering
\includegraphics[height=2.5cm]{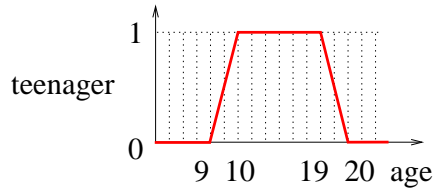} 
\caption{Teenager credibility.}
\label{fig:teenager_credibility}
\end{figure}

Fuzzy truth values are usually represented using continuous
functions. 
Fig. \ref{fig:teenager_credibility} shows an example in which 
the truth value function assigns the truth value of being 
{\it teenager} from the person's age value.

%
% The function is defined by means of points, as in a 2-D graphic. 
%
% \vspace{-0.5cm}
% \vspace{-1cm}

A function can be defined in several ways, but the easiest one is 
via a sequence of ordered pairs whose first element is the fact and the second element is the truth value 
assigned to that fact. 

Functions used to define the truth value of some group 
of facts are usually continuous and linear over intervals.
To define those functions there is no necessity to write down 
the value assigned to each element in their domains.
A better way to define them is by means of their
inflexion points, so function values for the elements between 
the inflexion points are determined by means of interpolation.

RFuzzy provides the syntax for defining functions by stretches. 
This syntax is shown in (\ref{eq-def-function}).
External brackets represent the Prolog list symbols and 
internal brackets represent cardinality in the formula notation. 
{\it arg1, ..., argN} should be ground terms (numbers) and {\it
truth\_val1, ..., truth\_valN} should be border truth values. The
truth value of the rest of the elements (apart from the border
elements) is obtained by interpolation.
\begin{equation}
pred\ \textbf{:\#\ }([ (arg1, truth\_val1),\ (arg2, truth\_val2)\ [,\ (arg3, truth\_val3)\ ]^* ])\ .
\label{eq-def-function}
\end{equation}

%% \medskip
%% \noindent
%% {\it Rfuzzy credibility function example}
\begin{verbatim}
:- set_prop teenager/1 => people_age/1.
:- default(teenager/1, 0).
teenager :# ([ (9, 0), (10, 1), (19, 1), (20, 0) ]) .
\end{verbatim}

%%%%%%%%%%%%   RULE TRUTH VALUE %%%%%%%%%%%%%%%%%%%%%%%%%%%%%%%%%%
\subsection{Rule truth value}
\label{rule-truth-value}

A tool which only allows the user to define truth values through 
functions and facts leaks on allowing him to combine those 
truth values for representing more complex situations.
A rule is the perfect tool to combine the truth values
of facts, functions, and other rules.

Rules allow the user to combine truth values in the correct way
(by means of aggregation operators, 
like {\it maximum} or {\it product}).
% and, if she wants to do it, to give that rules a credibility.
Besides this combination truth value for the body of the rule, 
the rule can be given an overall credibility truth value.

Credibility is used to 
express how much we trust the rule we write.
Suppose a small weather example in which we have the rule
{\it it will rain if it is cloudy and it is hot}. 
As it might rain but it might not, 
we can assign the rule a credibility of 0.7. 
As expected, the truth value obtained from the body is combined with
the credibility value of the rule to obtain a final truth value.

{\it Rfuzzy} offers the user a concrete syntax to define combinations 
of truth values by means of aggregation operations, and assign to 
that rules a credibility. 
This syntax extension is defined in (\ref{eq-def-predicate}).
Indeed, the user can choose two aggregation 
operators \footnote{Aggregation operators available are: 
{\it min} for minimum, {\it max} for maximum,
{\it prod} for the product, {\it luka} for the Lukasiewicz operator,
{\it dprod} for the inverse product, {\it dluka} for the inverse
Lukasiewicz operator and {\it complement}.}:
{\it op2} for combining the truth values of the subgoals of 
the rule's body and {\it op1} for combining the previous result 
with the credibility of the rule.

% \begin{equation}
\begin{eqnarray}
pred( arg1\ [,\ arg2]^*\ )\ [\ \textbf{cred\ (}op1, value1\textbf{)}\ ]^{0,1} \textbf{:} \thicksim 
op2~ pred1(arg1\ [,\ arg2]^*\ ) \nonumber \\
\ [,\ pred2(arg1\ [,\ arg2]^*\ ) ]\ .\ 
\label{eq-def-predicate}
\end{eqnarray}
% \end{equation}

The following examples show its usage. The second one uses 
the operator {\it prod} for aggregating truth values of the subgoals
of the body and the operator {\it min} to aggregate the result with the 
credibility of the rule, 0.8. 
As can be deduced from syntax and examples, 
{\it cred} and {\it :$\thicksim$} are reserved words.

\begin{verbatim}
tempting_restaurant(R) :~ prod low_distance(R), cheap(R), 
                               traditional(R).
\end{verbatim}
\begin{verbatim}
good_player(J) cred (min,0.8) :~ prod swift(J), tall(J), 
                                      experience(J).
\end{verbatim}

%%%%%%%%%%%%    GENERAL DEFAULT VALUE   %%%%%%%%%%%%%%%%%%%%%%%%%%%
%%%%%%%%%%%%    CONDITIONED DEFAULT VALUE   %%%%%%%%%%%%%%%%%%%%%%%
\subsection{General and Conditioned Default Truth Values}
\label{general-and-conditioned-default-truth-values}

Unfortunately, information provided by the user is not complete in
general. So there are many cases in which we have no information about
the truth value of an individual or a set of them. Nevertheless, it is
interesting not to stop a complex query evaluation just because we
have no information about one or more subgoals if we can use a
reasonable approximation. The solution to this problem is using
default truth values for these cases. The RFuzzy extension to define a
default truth value for a predicate when applied to individuals for
which the user has not defined an explicit truth value is named {\it
general default truth value}.  

{\it Conditioned default truth value} is used when the default truth value 
only applies to a subset of the function's domain.
This subset is defined by a membership predicate which is true 
only when an individual belongs to the subset.
The membership predicate ( {\it membership\_predicate/ar} ) 
and the predicate to which it is applied ( {\it pred/ar} ) 
need to be have the same arity ( {\it ar} ). 
If not, an error message will be shown at compilation time.

The syntax for defining a general default truth value 
is shown in (\ref{eq-def-defaults}),
and the syntax for assigning a conditioned default truth value  
is shown in (\ref{eq-def-default-with-conditions}).
{\it pred/ar} is in both cases the predicate to which we are
defining default values.
As expected, when defining the three cases (explicit, conditioned
and default truth value) only one will be given back when doing
a query. The precedence when looking for the truth value 
goes from the most concrete to the least one.
\begin{equation}
\textbf{:- default(}pred\textbf{/}ar,\ truth\_value\textbf{)\ .}
\label{eq-def-defaults}
\end{equation}
\begin{equation}
  \textbf{:- default(}pred\textbf{/}ar,\ truth\_value \textbf{) =\textgreater}\ membership\_predicate\textbf{/}ar\textbf{.}
\label{eq-def-default-with-conditions}
\end{equation}

The example below shows how to assign a default truth value of 0.5 
to all cars that do not have an explicit truth value nor have
a default conditioned truth value. 
Besides, it shows how to assign a conditioned default truth value 
to all cars belonging to a small subset and not having an explicit
truth value. 
This subset is determined by the membership predicate 
{\it expensive\_type/1}, 
and default truth value for its elements is 0.9.
So {\it lamborghini\_urraco} is an {\it
expensive\_car} with truth value 0.9 but {\it vw\_caddy} is an {\it
expensive\_car} with truth value 0.5. Both values are default
approximations because we have no direct declaration (as for {\it
alfa\_romeo\_gt} that is an {\it expensive\_car} with a truth value
0.6 as we show above).

\begin{verbatim}
:- set_prop expensive_car/1 => car/1.
:- default(expensive_car/1, 0.9) => expensive_type/1.
:- default(expensive_car/1, 0.5).

expensive_type(lamborghini_urraco).
expensive_type(aston_martin_bulldog).
\end{verbatim}

%%%%%%%%%%%%    QUERYING THE SYSTEM    %%%%%%%%%%%%%%%%%%%%%%%
\subsection{Doing queries with truth values}
\label{doing-queries-with-truth-values}

Indeed the program has to be run. 
When compiling, {\it Rfuzzy\ } adds a new argument to the 
arguments list of each fuzzy predicate.
This argument serves for querying about the predicate truth value.
It can be seen as syntactic sugar, 
as truth value is not part of the predicate arguments, 
but metadata information. 

Truth value argument is added to the predicates in a uniform way:
it is always a new argument at the end of the arguments list of the predicate.
In the previous example we wrote {\it expensive\_car/1}, so to query 
the system we have to give the predicate two arguments instead of only one
where the second one will represent the query's truth value.
This can be seen in the first example of subsection \ref{constructive-answers}.

%%%%%%%%%%%%    CONSTRUCTIVE ANSWERS   %%%%%%%%%%%%%%%%%%%%%%%
\subsection{Constructive Answers}
\label{constructive-answers}

A fuzzy tool
should be able to provide constructive answers for queries. The
regular (easy) questions are asking for the truth value of an
element. For example, how expensive is an {\it alfa\_romeo\_gt}?
\begin{verbatim}
?- expensive_car(alfa_romeo_gt,V).
V = 0.6 ? ;
no
\end{verbatim}
But the really interesting queries are the ones that ask for values
that satisfy constraints over the truth value. For example, which
cars are very expensive?. RFuzzy provides this constructive
functionality.
\begin{verbatim}
?- expensive_car(X,V), V > 0.8.
V = 0.9, X = aston_martin_bulldog ? ;
V = 0.9, X = lamborghini_urraco ? ;
no
\end{verbatim}

The RFuzzy package implements a meta-translation of the RFuzzy syntax
to ISO Prolog, via CLP(R), this is the reason for its
constructivity.

%%%%%%%%%%%%%%%%%%%%%%%%%%%%%%%%%%%%%%%%%%%%%%%%%%%%%%%%%%%%%%%%%%%
%%%%%%%%%%%%%%%    RFUZZY APPLICATIONS   %%%%%%%%%%%%%%%%%%%%%%%%%%
%%%%%%%%%%%%%%%%%%%%%%%%%%%%%%%%%%%%%%%%%%%%%%%%%%%%%%%%%%%%%%%%%%%

\section{Applications}

Rfuzzy is mainly suitable for expert systems applications.
As mentioned before, its main advantages in comparison to Fuzzy Prolog are 
its simpler syntax, the use of real numbers instead of intervals 
between them and the implicit handling of truth values.
Besides, it presents 
facts' truth values (explicit, default or conditioned default truth value), 
functions' truth values and rules (with or without credibility) 
which simplifies the user development process a lot.

Although a medical expert system development were the best example of  
using Rfuzzy, due to lack of space we prefer to show here one     
in which we decide which is the best restaurant for going out.       
% The reviewers noticed that there was no lack of space at all.
% So we should either present the medical expert system or 
% remove the apologetic remark.

\begin{verbatim}
:- module(restaurant,_,[rfuzzy, clpq]).

:- prop restaurant/1.

:- set_prop tempting_restaurant/1 => restaurant/1.
:- default(tempting_restaurant/1, 0.1).
tempting_restaurant(R) :~ prod low_distance(R), cheap(R), 
                               traditional(R).

restaurant(kenzo).
restaurant(burger_king).
restaurant(pizza_jardin).
restaurant(subway).
restaurant(derroscas).
restaurant(il_tempietto).
restaurant(kono_pizza).
restaurant(paellador).
restaurant(tapasbar).

crisp_distance(kenzo, 50).
crisp_distance(burguer_king, 100).
crisp_distance(pizza_jardin, 70).
crisp_distance(subway, 85).
crisp_distance(derroscas, 120).
crisp_distance(il_tempietto, 150).
crisp_distance(kono_pizza, 65).
crisp_distance(paellador, 55).
crisp_distance(tapasbar, 40).

low_distance(R, TV) :- crisp_distance(R, D), 
                       low_distance_aux(D, TV).

:- set_prop low_distance_aux/1 => distance/1.
:- default(low_distance_aux/1, 0).
low_distance_aux :# ([ (0, 1), (50, 0.9), (100, 0.8), (200, 0.6), 
                       (300, 0.5), (500, 0.4), (1000, 0.1), 
                       (2000, 0) ]).

very_low_distance(X) :- crisp_distance(X, D), D < 100.

:- set_prop cheap/1 => restaurant/1.
:- default(cheap/1, 0.2) => very_low_distance/1.
:- default(cheap/1, 0.5).

cheap(kenzo) value 0.2.
cheap(el_rincon) value 1.
cheap(el_reventaero) value 1.

:- set_prop traditional/1 => restaurant/1.
:- default(traditional/1, 0.8) => very_low_distance/1.
:- default(traditional/1, 1).

traditional(kenzo) value 0.5.
traditional(el_reventaero) value 0.87.

distance(0). 
distance(X) :- distance(Y), number(Y),  
	X .=. Y + 1,
	(   X < 5000 ; ( X >= 5000, !, fail) ).
\end{verbatim}

In the example we can see that we know the distance to all the 
restaurants in a crisp way. This crisp value is translated by means of
{\it low\_distance} and {\it low\_distance\_aux} into a fuzzy one 
which is used into {\it tempting\_restaurant} to determine its 
truth value. This allows us to ask which is the truth value of each 
tempting restaurant, which restaurant is a tempting 
restaurant with a truth value of, for example, 0.7 or list all 
tempting restaurants with their truth values.

The Rfuzzy module with installation instructions and examples can be
downloaded from \url{http://babel.ls.fi.upm.es/software/rfuzzy/}.

%%%%%%%%%%%%%%%%%%%%%%%%%%%%%%%%%%%%%%%%%%%%%%%%%%%%%%%%%%%%%%%%%%%
%%%%%%%%%%%%    IMPLEMENTATION DETAILS   %%%%%%%%%%%%%%%%%%%%%%%%%%
%%%%%%%%%%%%%%%%%%%%%%%%%%%%%%%%%%%%%%%%%%%%%%%%%%%%%%%%%%%%%%%%%%%

\section{Implementation details}

RFuzzy is a logic programming language that is able to model 
all the extensions that are described in section
\ref{rfuzzy:syntax}. 
It is implemented as a Ciao Prolog \cite{ciao-prolog:site} 
package because Ciao Prolog offers the possibility of dealing 
with a higher order compilation through the implementation 
of Ciao packages.
Those packages serve as input for the 
{\it``Ciao System Preprocessor''} (CiaoPP) \cite{ciaopp-manual}, 
a tool able to perform source-to-source transformations.

The reason beyond the implementation of {\it Rfuzzy } as a Ciao 
Prolog package is that 
the resultant code has to deal with two kinds of queries:
\begin{itemize}
\item queries in which the user asks for the truth value of an individual, and
\item queries in which the user asks for an individual with a concrete truth value.
\end{itemize}

As can be seen in the following example, 
this is not an easy task.
\begin{verbatim}
?- A is 1, B is 2, C is A + B.

A = 1, B = 2, C = 3 ? .
yes
?- C is 3, C is A + B.
{ERROR: illegal arithmetic expression}
{ERROR: illegal arithmetic expression}
no
?- 
\end{verbatim}

Formula {\it C is A + B} only works if variables A and B are bound.
Almost all predicates that are problematic with non-bound 
variables have inside comparisons and/or assignments. 
This aims us to translate Rfuzzy programs into CLP(${\cal R}$) 
programs. 
CLP(${\cal R}$) is a Ciao Prolog Package which translates real
number operations into constraints applied to the variables 
involved in those operations. 

\begin{figure}
\centering
\includegraphics[height=1.5cm]{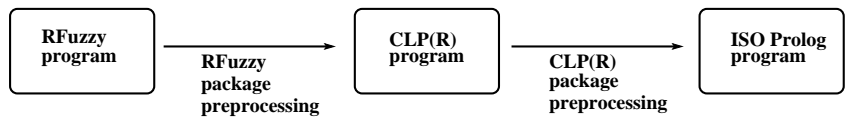} 
\caption{Rfuzzy architecture.}
\label{fig:rfuzzy_architecture}
\end{figure}

Taking advantage of Rfuzzy and CLP(${\cal R}$) transformations, 
our tool compiles Rfuzzy programs into ISO Prolog programs,
so the interpreter is able to work with them as it normally does.
As a result, the global compilation process has two 
preprocessor steps:
(1) the Rfuzzy program is translated into CLP(${\cal R}$) 
constraints by means of the Rfuzzy package and 
(2) those constraints are translated into ISO Prolog by using 
the CLP(${\cal R}$) package. 
Fig. \ref{fig:rfuzzy_architecture} shows the whole process.
%
%\vspace{-0.3cm}
%\vspace{-1cm}

In the following example the predicate {\it tempting\_restaurant}
is translated from Rfuzzy syntax into ISO Prolog syntax.
In the first step, the Rfuzzy package inserts truth value variables, 
the {\it inject} metapredicate call
(one of its arguments is the aggregation operator to be used, 
{\it prod}) 
and inserts Rfuzzy comparisons to take care at execution time 
that the rule's truth value is always between zero and one.
In the second step, CLP(${\cal R}$) converts comparisons into 
constraints (via predicate calls).

\begin{verbatim}
% Rfuzzy program
tempting_restaurant(R) :~ prod low_distance(R), cheap(R), 
                               traditional(R).
\end{verbatim}

\begin{verbatim}
% CLP(R) program
rfuzzy_rule_tempting_restaurant(R,_1) :-
        low_distance(R,_2),
        cheap(R,_3),
        traditional(R,_4),
        inject([_2,_3,_4],prod,_1),
        _1 .>=. 0, 
        _1 .=<. 1.
\end{verbatim}

\begin{verbatim}
% ISO Prolog program
rfuzzy_rule_tempting_restaurant(R,_1) :-
        low_distance(R,_2),
        cheap(R,_3),
        traditional(R,_4),
        inject([_2,_3,_4],prod,_1),
        solve_generic_1(le,0,_1,-1),
        solve_generic_1(le,-1,_1,1) .
\end{verbatim}

Internally, Rfuzzy package unifies and translates all the information
given by the user to each predicate 
(Types, default values with and without condition, truth values
defined in facts and rules with and without credibility) 
into a single predicate body. 
A simplified version of the skeleton used for that predicate is shown
below.

\medskip
\noindent
{\it Rfuzzy package simplified skeleton }
\begin{verbatim}

Main :- Types, ( Normal ; Default)
 
Normal :- ( Fact ; 
            (\+(Fact_Aux), Function) ;  
            (\+(Fact_Aux), \+(Function_Aux), Rule) 
          )

Default :- \+(Fact_Aux), \+(Function_Aux), \+(Rule_Aux), 
             ( Cl_With_Cond ; 
               (\+(Cl_With_Cond_Aux), Cl_With_No_Cond) 
             )

\end{verbatim}

The skeleton has three different parts: the one which takes care of
allowing only queries or answers with the expected individuals, 
the one which looks for a concrete truth value (it can be defined
by means of a fact, a function or a rule) and the one which 
looks for a default truth value (conditioned or not).
Predicates ending in \_aux do not take care on the 
truth value argument. 

The first part is obtained from the type definitions (see \ref{type-definition}),
translating all types into a predicate which is called at 
first (Types) so we assure we only work with the expected 
individuals. 

The second part looks for a concrete value whose arguments have to unify 
with the parameters the user has given. 
Precedence when looking for it is: 
\begin{enumerate}
\item A fact (see subsection \ref{fact-truth-value})
\item A function (see subsection \ref{functional-truth-value})
\item A rule (see subsection \ref{rule-truth-value})
\end{enumerate}

The third part is only called when the second one (searching for a 
truth value) fails, and looks for a conditioned or default truth value.

%%%%%%%%%%%%%%%%%%%%%%%%%%%%%%%%%%%%%%%%%%%%%%%%%%%%%%%%%%%%%%%%%%%
%%%%%%%%%%%%    CONCLUSIONS    %%%%%%%%%%%%%%%%%%%%%%%%%%%%%%%%%%%%
%%%%%%%%%%%%%%%%%%%%%%%%%%%%%%%%%%%%%%%%%%%%%%%%%%%%%%%%%%%%%%%%%%%

\section{Conclusions and Current Work}

{\it Rfuzzy} offers to the users a new framework to represent 
fuzzy problems over real numbers. 
{\it Fuzzy Prolog} 
\cite{Vaucheret_ICLP02,Vaucheret_LPAR02,Susana_FSS04}
is an existing framework for dealing with Fuzzy problems
representation.  
Main {\it Rfuzzy} advantages over {\it Fuzzy Prolog} are 
a simpler syntax and the elimination of answers with constraints,
and {\it Rfuzzy} is one of the first tools modelling 
multi-adjoint logic, as explained in subsection 
\ref{subsec:motivation}.

{\it Rfuzzy} syntax is simpler that {\it Fuzzy Prolog} syntax.
Its fuzzy values are simple real numbers instead of 
intervals between real numbers, and it hides the
management of truth value variables.
As normal fuzzy problems do not use intervals to represent 
fuzziness and do not need to code an uncommon behaviour of 
fuzzy variables, this syntax reduction is an advantage. 
Programs written in {\it Rfuzzy} syntax are more legible
and more easy to understand than {\it Fuzzy Prolog} programs.

{\it Fuzzy Prolog} answers to user queries are difficult to 
understand due to the existence of constraints. 
As normal replies to final users are ground terms, 
the programmer has to code by hand how to reach them.
To eliminate those constraints and answer queries with 
ground terms the programmer tries to substitute variables 
with ground terms until one makes true all of them.
{\it Rfuzzy} offers a powerful tool to deal with this 
task: {\it Type definition}. 
{\it Type definition} (see subsection \ref{type-definition})
allows the user to define which terms are suitable for 
being substituted into a variable, so she does not have to 
code this behaviour again. 
Besides, the elimination of answers with constraints provides 
more human readable answers and more easy to test programs
(because answers we test do not have constraints, 
just ground terms). 

There is also an extension to introduce default truth values. 
As world information is sometimes incomplete, {\it Rfuzzy}
offers to the user the possibility to define default truth 
values and default conditioned truth values (see subsection 
\ref{general-and-conditioned-default-truth-values}).
This allows us to make inference with default truth values
when we do not know anything about the truth of some fact.

Extensions added to {\it Prolog} by {\it Rfuzzy} are:
Types, default truth values (conditioned or not),
assignment of truth values to individuals by means of facts, 
functions or rules, 
and assignment of credibility to the rules.

Besides, the possibility to provide constructive answers 
to the queries increase its usage, as can be seen in 
subsection \ref{constructive-answers}. 

There are countless applications and research lines which can benefit 
from the advantages of using the fuzzy representations offered
by Rfuzzy. Some examples are: 
Search Engines, Knowledge Extraction (from databases, ontologies, etc.), 
Semantic Web, Business Rules, and
Coding Rules (where the violation of one rule can be given a truth value).

Current work on Rfuzzy tries to apply constructive negation to the
engine.  RFuzzy needs to define types in a constructive way 
(by means of predicates that are able to generate all their 
individuals by backtracking) 
so we cannot use constraints.
Future research will be done in this line for widening the 
definition of types.

%%%%%%%%%%%%%%%%%%%%%%%%%%%%%%%%%%%%%%%%%%%%%%%%%%%%%%%%%%%%%%%%%%%
%%%%%%%%%%%%    BIBLIOGRAPHY   %%%%%%%%%%%%%%%%%%%%%%%%%%%%%%%%%%%%
%%%%%%%%%%%%%%%%%%%%%%%%%%%%%%%%%%%%%%%%%%%%%%%%%%%%%%%%%%%%%%%%%%%

% \bibliographystyle{plain}  %alpha yplain
% \bibliographystyle{abbrv}  %alpha yplain
% \bibliography{bib_rfuzzy}

\begin{thebibliography}{10}

\bibitem{Abietar07}
J.M. Abietar, P.J. Morcillo, and G.~Moreno.
\newblock Designing a software tool for fuzzy logic programming.
\newblock In T.E. Simos and G.~Maroulis, editors, {\em Proc. of the Int. Conf.
  of Computational Methods in Sciences and Engineering. ICCMSE'07}, volume~2 of
  {\em Computation in Mordern Science and Engineering}, pages 1117--1120.
  American Institute of Physics, 2007.
\newblock Distributed by Springer.

\bibitem{fril}
J.~F. Baldwin, T.~P. Martin, and B.~W. Pilsworth.
\newblock {\em Fril: Fuzzy and Evidential Reasoning in Artificial
  Intelligence}.
\newblock John Wiley \& Sons, 1995.

\bibitem{Bistarelli01}
S.~Bistarelli, U.~Montanari, and F.~Rossi.
\newblock Semiring-based constraint {L}ogic {P}rogramming: syntax and
  semantics.
\newblock In {\em ACM TOPLAS}, volume~23, pages 1--29, 2001.

\bibitem{ciaopp-manual}
F.~Bueno, P.~L\'{o}pez-Garc\'{\i}a, G.~Puebla, and M.~Hermenegildo.
\newblock {T}he {Ciao} {P}rolog {P}reprocessor.
\newblock Technical Report CLIP8/95.0.7.20, Technical University of Madrid
  (UPM), Facultad de Inform\'atica, 28660 Boadilla del Monte, Madrid, Spain,
  1999.

\bibitem{Cao00}
T.H. Cao.
\newblock Annotated fuzzy logic programs.
\newblock {\em Fuzzy Sets and Systems}, 113(2):277--298, 2000.

\bibitem{Dubois91}
D.~Dubois, J.~Lang, and H.~Prade.
\newblock Towards possibilistic logic programming.
\newblock In {\em Proc. of ICLP-91}, pages 581--595. MIT Press, 1991.

\bibitem{Fitting91}
M.~Fitting.
\newblock Bilattices and the semantics of logic programming.
\newblock {\em Journal of {L}ogic Programmig}, 11:91--116, 1991.

\bibitem{Fuhr00}
N.~Fuhr.
\newblock Probabilistic datalog: Implementing logical information retrieval for
  advanced applications.
\newblock {\em Journal of the American Society for Information Science},
  51(2):95--110, 2000.

\bibitem{Susana_FSS04}
S.~Guadarrama, S.~Munoz-Hernandez, and C.~Vaucheret.
\newblock Fuzzy {P}rolog: A new approach using soft constraints propagation.
\newblock {\em Fuzzy Sets and Systems, FSS}, 144(1):127--150, 2004.
\newblock ISSN 0165-0114.

\bibitem{ishizuka85prologelf}
M.~Ishizuka and N.~Kanai.
\newblock Prolog-{ELF} incorporating fuzzy {L}ogic.
\newblock In {\em {IJCAI}}, pages 701--703, 1985.

\bibitem{Kifer88}
M.~Kifer and Ai~Li.
\newblock On the semantics of rule-based expert systems with uncertainty.
\newblock In {\em Proc. of ICDT-88}, number 326 in LNCS, pages 102--117, 1988.

\bibitem{Kifer92}
M.~Kifer and V.S. Subrahmanian.
\newblock Theory of generalized annotated logic programming and its
  applications.
\newblock {\em Journal of {L}ogic {P}rogramming}, 12:335--367, 1992.

\bibitem{klawonn94lukasiewicz}
F.~Klawonn and R.~Kruse.
\newblock A {{\L{}}}ukasiewicz logic based {P}rolog.
\newblock {\em Mathware \& Soft Computing}, 1(1):5--29, 1994.

\bibitem{Norms00}
E.P. Klement, R.~Mesiar, and E.~Pap.
\newblock Triangular norms.
\newblock Kluwer Academic Publishers.

\bibitem{Lakshmanan94}
L.~Lakshmanan.
\newblock An epistemic foundation for logic programming with uncertainty.
\newblock {\em LNCS}, 880:89--100, 1994.

\bibitem{LakshmananShiri94}
L.~Lakshmanan and N.~Shiri.
\newblock Probabilistic deductive databases.
\newblock {\em Int. {L}ogic {P}rogramming Symposium}, pages 254--268, 1994.

\bibitem{Lakshmanan01}
L.~Lakshmanan and N.~Shiri.
\newblock A parametric approach to deductive databases with uncertainty.
\newblock {\em IEEE Transactions on Knowledge and Data Engineering},
  13(4):554--570, 2001.

\bibitem{lee1}
R.~C.~T. Lee.
\newblock Fuzzy {L}ogic and the resolution principle.
\newblock {\em Journal of the Association for Computing Machinery},
  19(1):119--129, 1972.

\bibitem{li90fuzzy}
D.~Li and D.~Liu.
\newblock {\em A Fuzzy Prolog Database System}.
\newblock John Wiley \& Sons, New York, 1990.

\bibitem{Lukasiewicz01}
T.~Lukasiewicz.
\newblock Fixpoint characterizations for many-valued disjunctive logic programs
  with probabilistic semantics.
\newblock In {\em Proc. of LPNMR-01}, volume 2173, pages 336--350, 2001.

\bibitem{M_Adjoint_Completeness}
J.~Medina, M.~Ojeda-Aciego, and P.~Votjas.
\newblock A completeness theorem for multi-adjoint {L}ogic {P}rogramming.
\newblock In {\em International Fuzzy Systems Conference}, pages 1031--1034.
  IEEE, 2001.

\bibitem{M_Adjoint_Continuous}
J.~Medina, M.~Ojeda-Aciego, and P.~Votjas.
\newblock Multi-adjoint {L}ogic {P}rogramming with continuous semantics.
\newblock In {\em LPNMR}, volume 2173 of {\em LNCS}, pages 351--364, Boston, MA
  (USA), 2001. Springer-Verlag.

\bibitem{M_Adjoint_Procedural}
J.~Medina, M.~Ojeda-Aciego, and P.~Votjas.
\newblock A procedural semantics for multi-adjoint {L}ogic {P}rogramming.
\newblock In {\em EPIA}, volume 2258 of {\em LNCS}, pages 290--297, Boston, MA
  (USA), 2001. Springer-Verlag.

\bibitem{Susana_fuzzyneg_AGP02}
S.~Munoz-Hernandez, C.~Vaucheret, and S.~Guadarrama.
\newblock Combining crisp and fuzzy {L}ogic in a prolog compiler.
\newblock In J.~J. Moreno-Navarro and J.~Mari{\~n}o, editors, {\em Joint Conf.
  on Declarative {P}rogramming: APPIA-GULP-PRODE 2002}, pages 23--38, Madrid,
  Spain, September 2002.

\bibitem{Ng91}
R.~Ng and V.S. Subrahmanian.
\newblock Stable model semantics for probabilistic deductive databases.
\newblock In {\em Proc. of ISMIS-91}, number 542 in LNCS, pages 163--171, 1991.

\bibitem{Ng93}
R.~Ng and V.S. Subrahmanian.
\newblock Probabilistic logic programming.
\newblock {\em Information and Computation}, 101(2):150--201, 1993.

\bibitem{Prad_Tri_Cal}
A.~Pradera, E.~Trillas, and T.~Calvo.
\newblock A general class of triangular norm-based aggregation operators:
  quasi-linear t-s operators.
\newblock {\em International Journal of Approximate Reasoning}, 30(1):57--72,
  2002.

\bibitem{Shen}
Z.~Shen, L.~Ding, and M.~Mukaidono.
\newblock Fuzzy resolution principle.
\newblock In {\em Proc. of 18th International Symposium on Multiple-valued
  {L}ogic}, volume~5, 1989.

\bibitem{Subrahmanian87}
V.S. Subrahmanian.
\newblock On the semantics of quantitative logic programs.
\newblock In {\em Proc. of 4th IEEE Symp. on {L}ogic {P}rogramming}, pages
  173--182. Computer Society Press, 1987.

\bibitem{ciao-prolog:site}
{The CLIP Lab}.
\newblock {The Ciao Prolog Development System WWW Site}, \\
  \texttt{http://www.clip.dia.fi.upm.es/Software/Ciao/}.

\bibitem{Tri_Cub_Cas}
E.~Trillas, S.~Cubillo, and J.~L. Castro.
\newblock Conjunction and disjunction on $([0,1],<=)$.
\newblock {\em Fuzzy Sets and Systems}, 72:155--165, 1995.

\bibitem{Emden86}
M.H. van Emden.
\newblock Quantitative duduction and its fixpoint theory.
\newblock {\em Journal of {L}ogic {P}rogramming}, 4(1):37--53, 1986.

\bibitem{Vaucheret_LPAR02}
C.~Vaucheret, S.~Guadarrama, and S.~Munoz-Hernandez.
\newblock Fuzzy prolog: A simple general implementation using clp(r).
\newblock In M.~Baaz and A.~Voronkov, editors, {\em {L}ogic for {P}rogramming,
  Artificial Intelligence, and Reasoning, LPAR 2002}, number 2514 in LNAI,
  pages 450--463, Tbilisi, Georgia, October 2002. Springer-Verlag.

\bibitem{Vaucheret_ICLP02}
C.~Vaucheret, S.~Guadarrama, and S.~Munoz-Hernandez.
\newblock Fuzzy prolog: A simple general implementation using clp(r).
\newblock In P.J. Stuckey, editor, {\em Int. Conf. in {L}ogic {P}rogramming,
  ICLP 2002}, number 2401 in LNCS, page 469, Copenhagen, Denmark, July/August
  2002. Springer-Verlag.

\bibitem{Vojt1}
P.~Vojtas.
\newblock Fuzzy logic programming.
\newblock {\em Fuzzy Sets and Systems}, 124(1):361--370, 2001.

\bibitem{Wagner97}
G.~Wagner.
\newblock A logical reconstruction of fuzzy inference in databases and logic
  programs.
\newblock In {\em Proc. of IFSA-97}, Prague, 1997.

\bibitem{Wagner98}
G.~Wagner.
\newblock Negation in fuzzy and possibilistic logic programs.
\newblock In {\em {L}ogic programming and Soft Computing}. Research Studies
  Press, 1998.

\bibitem{Shapiro83}
Ehud Y. and Shapiro.
\newblock {L}ogic programs with uncertainties: A tool for implementing
  rule-based systems.
\newblock In {\em IJCAI}, pages 529--532, 1983.

\end{thebibliography}

\end{document}